\documentclass[%
 reprint,
 amsmath,amssymb,
 aps,
]{revtex4-2}
\usepackage{ytableau}
\usepackage[hyperfootnotes=false]{hyperref}
\hypersetup{
 colorlinks=true,
 citecolor=blue,
 linkcolor=red,
 urlcolor=blue}
\usepackage{mathtools}
\usepackage{nccmath} 
\usepackage{xcolor}
\usepackage{caption}
\usepackage{subcaption}
\usepackage{graphicx,physics}
\usepackage{dcolumn}
\usepackage{bm}

\begin{document}

\preprint{APS/123-QED}

\title{Phenomenological Analysis of Triply Heavy Pentaquarks with configurations $q\Bar{q}QQQ$ and $qqQQ\Bar{Q}$.}

\author{Ankush Sharma, Alka Upadhyay}
 \affiliation{Department of Physics and Material Sciences, Thapar Institute of Engineering and Technology, Patiala, India, 147004}
 
\email{ankushsharma2540.as@gmail.com}

\date{\today}

\begin{abstract}
We carried out the systematic analysis of the $s$-wave triply heavy pentaquarks with possible configurations like $q\Bar{q}QQQ$ and $qqQQ\Bar{Q}$, ($q = u, d, s$ and $Q = c, b$ quarks). Special unitary representations are utilized to study the classification scheme for triply heavy configurations like $q\Bar{q}QQQ$ and $qqQQ\Bar{Q}$. We classified the $q\Bar{q}QQQ$-type pentaquarks into an octet and $qqQQ\Bar{Q}$-type pentaquarks into sextet configurations with the help of SU(3) flavor representation. Also, with the help of SU(2) spin representation, we studied the possible spin assignments ($\frac{1}{2}^-$, $\frac{3}{2}^-$, and $\frac{5}{2}^-$) for ground state triply heavy pentaquarks. Furthermore, we used the formalism of the extended Gursey-Radicati mass formula and effective mass scheme to estimate the masses of triply heavy pentaquarks. Additionally, we calculated the magnetic moment assignments using the effective mass and screened charge schemes. The predicted outcomes align well with the existing theoretical data and benefit future studies. Our work provides a comprehensive framework that combines the theoretical aspects of SU(3) and SU(2) symmetries with practical predictions for observables, offering a strong foundation for experimental verification. This integrated approach enhances our understanding of the complex interactions within triply heavy pentaquarks and underscores their potential role in probing deeper into the dynamics of the strong force. These findings are crucial for designing future high-energy experiments that directly observe these exotic states and confirm their properties, paving the way for new insights into quantum chromodynamics.
\end{abstract}

\maketitle

\section{Introduction}
In recent years, there has been notable growth in both experimental and theoretical research within the field of exotic hadrons. Experimental collaborations like LHCb identified many tetraquark and pentaquark structures with a statistical significance of more than 5 $\sigma$. Recently, LHCb collaboration observed two tetraquarks with a single heavy quark having quark contents $u\bar{d}c\bar{s}$ and $\bar{u}dc\bar{s}$. These tetraquark structures were observed with a statistical significance of 6.5 $\sigma$ and 8 $\sigma$, with mass and width are $M_{exp} = 2908\pm 11 \pm 20 MeV$ and $\Gamma_{exp}= 136 \pm 23 \pm 13  MeV$, respectively \cite{2900}. Also, a pentaquark state, $P_{\psi s}^\Lambda(4338)^0$ with quark content $udsc\bar{c}$ observed with a statistical significance of 15 standard deviations \cite{4338}. The mass and width of $P_{\psi s}^\Lambda(4338)^0$ state are measured to be $4338.2\pm0.7\pm0.4$ MeV and $7.0\pm1.2\pm1.3$ MeV, respectively. Moreover, in 2015, the LHCb collaboration identified two hidden-charm pentaquark structures known as $P_c(4380)$ and $P_c(4450)$. \cite{PhysRevLett.115.072001}. The masses of the two $P_c^+$ states are measured to be 
$4380 \pm 8 \pm 29$ MeV and $4449.8 \pm 1.7 \pm 2.5 $ MeV, with corresponding widths of $205 \pm 18 \pm 86 $ MeV and $39 \pm 5 \pm 19$ MeV. In 2019, a narrow pentaquark structure, $P_c(4312)^+$ was discovered with a statistical significance of $7.3\sigma$ via $\Lambda_b^0 \rightarrow J/\psi p K^-$ decays \cite{PhysRevLett.122.222001}. In 2020, the first evidence of a strange hidden-charm structure was observed from an amplitude analysis of $\Xi_b^- \rightarrow J/\psi \lambda K^-$ decays. The mass and width are  $4458.8 \pm 2.94_{-1:1}^{+4.7}$ MeV and $17.3 \pm 6.5_{-5.7}^{+8.0}$ MeV, respectively \cite{20211278}. These ongoing observations inspired theoreticians to examine the dynamics of exotic hadrons. Many theoretical frameworks has been propsed to study the pentaquark states since the discovery of $\Theta^+(1540)$ pentaquark state \cite{PhysRevC1,PhysRevC2,PhysRevC3,PhysRevC4,PhysRevC5,PhysRevC6,PhysRevC7}. This study focused on analyzing the masses and magnetic moments of triply heavy pentaquarks with configurations $q\Bar{q}QQQ$ and $qqQQ\Bar{Q}$, ($q = u, d, s$ and $Q = c, b$ quarks). Various theoretical models, such as QCD sum rules and the chromomagnetic interactions, have been suggested to investigate the dynamics of triply heavy pentaquarks \cite{EPJC1, EPJC2}. Understanding the properties and behavior of triply heavy pentaquarks is crucial for several reasons. Firstly, these exotic particles provide valuable insights into matter's fundamental forces and constituents. Secondly, studying triply heavy pentaquarks can expand our knowledge of the Standard Model of particle physics, potentially revealing new physics beyond the currently established framework. Triply heavy pentaquarks represent an unexplored region of the particle spectrum, offering opportunities to discover new exotic states and phenomena. Studying their masses, decays, and interactions contributes to mapping out the comprehensive spectrum of hadronic matter. We used the formalism of the Gursey-Radicati mass formula and effective mass scheme to calculate the masses of ground-state triply heavy pentaquarks for each possible spin assignment. By integrating the complex interaction of quantum properties such as spin, isospin, and flavor quantum numbers with the dynamics of particle interactions, the GR mass formula offers a comprehensive understanding of mass generation mechanisms \cite{FG}. The formalism of the Gursey-Radicati mass formula has proven to be an effective method for calculating the masses of exotic hadrons in recent times \cite{Santo, HOLMA, Sharma_2023, Ankush}. Furthermore, pentaquarks are exotic hadrons, which consist of four quarks and an antiquark, and exploring their internal composition is captivating. Understanding their magnetic moment assignments is crucial for gaining insight into their internal structure and the behavior of constituent quarks. Magnetic moments are sensitive tools used to investigate the distribution of charge and current within pentaquarks and can be utilized to verify the predictions of quantum chromodynamics (QCD). Using effective mass and screened charge schemes, we estimated the magnetic moments of triply heavy pentaquarks for both $q\Bar{q}QQQ$ and $qqQQ\Bar{Q}$ type configurations. The effective mass scheme enables researchers to explore various aspects of triply heavy pentaquarks, such as their mass spectra, magnetic moments, and other observables, within a unified theoretical framework. Also, applying the screened charge scheme to triply heavy pentaquarks opens up new avenues for exploring the dynamics of quark-gluon interactions and advancing our knowledge of the strong nuclear force at the most fundamental level.\\

This work is organized as follows: section II consists of the theoretical formalism, which briefly describes the classification scheme for triply heavy pentaquarks using the special unitary representations and the Young Tableau technique, corresponding spin-flavor wave-functions, etc., the extension of the GR mass formula, and the effective mass scheme to calculate the masses of triply heavy pentaquarks, and the screened charge schemes for magnetic moment assignments. Section III includes the results and discussion about $qqQQ\Bar{Q}$ and $q\Bar{q}QQQ$ - type pentaquarks, and section IV concludes the article.    

\section{Theoretical Formalism}
Studying the properties of triply heavy pentaquarks is interesting for several reasons. Firstly, they offer a unique opportunity to explore the fundamental interactions of quarks under extreme conditions, providing insights into the strong force that binds them together. Additionally, triply heavy pentaquarks can exhibit exotic behaviors due to their complex quark composition, which may challenge our current understanding of particle physics. Furthermore, these particles could shed light on the formation and structure of hadrons, enriching our knowledge of the quark-gluon dynamics within them. In this work, we calculated the masses and magnetic moments of $qqQQ\Bar{Q}$ and $q\Bar{q}QQQ$ - type triply heavy pentaquarks by classifying them into allowed spin-flavor multiplets of special unitary representation. Furthermore, an extended version of the Gursey-Radicati mass formula, effective mass scheme, and screened charge scheme have been used to estimate their masses and magnetic moments. Detailed overviews are written as follows:

\subsection{Classification scheme for Triply heavy Pentaquarks}
We employed special unitary representations to examine the classification system for triply heavy pentaquarks. With the help of SU(3) flavor representation, every quark is designated by the fundamental 3 representation, while the antiquark is assigned the $\Bar{3}$ representation. Therefore, the classification scheme for $qqQQ\Bar{Q}$ - type pentaquarks is:
\begin{equation}
    [3] \otimes [3] = [\Bar{3}] \oplus [6]
    \label{1}
\end{equation}
and for $q\Bar{q}QQQ$ - type pentaquarks is:
\begin{equation}
    [3] \otimes [\Bar{3}] = 1 \oplus [8]
    \label{2}
\end{equation}
By using equation \eqref{1} and \eqref{2}, we classified the  $q\Bar{q}QQQ$ type pentaquarks into an octet and $qqQQ\Bar{Q}$ into sextet configurations. Moreover, with the help of SU(2) spin representation, assigning both quarks and antiquarks to fundamental '2' representations, we have:
\begin{align}
    [2] \otimes [2] \otimes [2] \otimes [2] \otimes
[2] =  [6]  \oplus 4[4] \oplus 5[2]
\label{3}
\end{align}
The configurations represented by [6], [4], and [2] correspond to spin values of $s$ = 5/2, 3/2, and 1/2, respectively. By using equations \eqref{1}, \eqref{2}, and \eqref{3}, we estimated the mass spectra and magnetic moments of triply heavy pentaquarks. The corresponding Young Tableau representation can be shown as:\\
\begin{table}[h]
\centering
    \caption{Young Tableau representation for Pentaquarks  \cite{young}.}
    \begin{tabular}{ccc}
     \begin{ytableau}
     \none &
\end{ytableau} $\otimes$
\begin{ytableau}
     \none &
\end{ytableau} $\otimes$
\begin{ytableau}
     \none &
\end{ytableau}  $\otimes$
 \begin{ytableau}
     \none &
\end{ytableau} $\otimes$
\begin{ytableau}
     \none &
\end{ytableau} = \nonumber \\ 
\\
\begin{ytableau}
    \none & & & & &
\end{ytableau} $\oplus$ 
4\begin{ytableau}
    \none & & & & \\
    \none &
\end{ytableau} $\oplus$
5\begin{ytableau}
    \none & & & \\
    \none & &
\end{ytableau}\\ 
\label{YT1}
    \end{tabular}
\end{table}
\\
The corresponding Young Tableau representations for spin-5/2, 3/2, and 1/2 are reported in Ref. \cite {young}. There are ten spin symmetries for pentaquarks, which were shown with the help of Young Tableau representation. The spin wave functions for ten symmetries for pentaquarks are defined as \cite{magneticmom}: 
\begin{widetext}
\begin{align}
     \chi_1 =& \ket{\uparrow\uparrow\uparrow\uparrow\uparrow} \\
     \chi_2 =& \frac{2}{\sqrt{5}}\ket{\uparrow\uparrow\uparrow\uparrow\downarrow - \frac{\sqrt{5}}{10}(\uparrow\uparrow\uparrow\downarrow\uparrow + \uparrow\uparrow\downarrow\uparrow\uparrow + \uparrow\downarrow\uparrow\uparrow\uparrow + \downarrow\uparrow\uparrow\uparrow\uparrow)}\\
    \chi_3 =& \frac{\sqrt{3}}{2}\ket{\uparrow\uparrow\uparrow\downarrow\uparrow - \frac{1}{2\sqrt{3}}(\uparrow\uparrow\downarrow\uparrow\uparrow + \uparrow\downarrow\uparrow\uparrow\uparrow + \downarrow\uparrow\uparrow\uparrow\uparrow)}\\
   \chi_4 =& \frac{1}{\sqrt{6}}\ket{(2\uparrow\uparrow\downarrow\uparrow\uparrow - \uparrow\downarrow\uparrow\uparrow\uparrow - \downarrow\uparrow\uparrow\uparrow\uparrow)}\\ 
   \chi_5 =& \frac{1}{\sqrt{2}}\ket{(\uparrow\downarrow\uparrow\uparrow\uparrow - \downarrow\uparrow\uparrow\uparrow\uparrow)}\\
   \chi_6 =& \frac{1}{3\sqrt{2}}\ket{(\uparrow\downarrow\downarrow\uparrow\uparrow + \downarrow\uparrow\downarrow\uparrow\uparrow + \downarrow\downarrow\uparrow\uparrow\uparrow - \uparrow\uparrow\downarrow\downarrow\uparrow - \uparrow\downarrow\uparrow\downarrow\uparrow 
   - \downarrow\uparrow\uparrow\downarrow\uparrow - \uparrow\uparrow\downarrow\uparrow\downarrow - \uparrow\downarrow\uparrow\uparrow\downarrow - \downarrow\uparrow\uparrow\uparrow\downarrow) + \frac{1}{\sqrt{2}} \uparrow\uparrow\uparrow\downarrow\downarrow}\\
   \chi_7 =& \frac{1}{3} \ket{(2\uparrow\uparrow\downarrow\uparrow\downarrow - \uparrow\downarrow\uparrow\uparrow\downarrow - \downarrow\uparrow\uparrow\uparrow\downarrow - \uparrow\uparrow\downarrow\downarrow\uparrow + \downarrow\downarrow\uparrow\uparrow\uparrow) + \frac{1}{6}(\uparrow\downarrow\uparrow\downarrow\uparrow - \uparrow\downarrow\downarrow\uparrow\uparrow - \downarrow\uparrow\downarrow\uparrow\uparrow + \downarrow\uparrow\uparrow\downarrow\uparrow)} \\
   \chi_8 =& \frac{1}{\sqrt{3}} \ket{(\uparrow\downarrow\uparrow\uparrow\downarrow - \downarrow\uparrow\uparrow\uparrow\downarrow) -\frac{1}{2\sqrt{3}}(\uparrow\downarrow\uparrow\downarrow\uparrow + \uparrow\downarrow\downarrow\uparrow\uparrow - \downarrow\uparrow\downarrow\uparrow\uparrow - \downarrow\uparrow\uparrow\downarrow\uparrow)}\\
 \chi_9 =& \frac{1}{\sqrt{3}}\ket{(\uparrow\uparrow\downarrow\downarrow\uparrow + \downarrow\downarrow\uparrow\uparrow\uparrow\uparrow) -\frac{1}{2\sqrt{3}}(\uparrow\downarrow\uparrow\downarrow\uparrow + \uparrow\downarrow\downarrow\uparrow\uparrow + \downarrow\uparrow\downarrow\uparrow\uparrow + \downarrow\uparrow\uparrow\downarrow\uparrow )}\\
 \chi_{10} =& \frac{1}{2}\ket{(\uparrow\downarrow\uparrow\downarrow\uparrow - \uparrow\downarrow\downarrow\uparrow\uparrow + \downarrow\uparrow\downarrow\uparrow\uparrow - \downarrow\uparrow\uparrow\downarrow\uparrow)}
\end{align}
\end{widetext}
The flavor wavefunction for a few of the triply heavy pentaquark states is also defined, and for $u\Bar{u}ccc$ state, it can be written as:
\begin{equation}
    \phi_{1} = \frac{1}{\sqrt{20}}\ket{[cccu + uccc + cucc + ccuc]\Bar{u}}
\end{equation}
\begin{widetext}
Similarly, we can define the flavor wavefunction for $u\Bar{u}bbb$ state. Moreover, for $uucc\Bar{c}$ state, flavor wavefunction is:
\begin{align}
  \phi_{4} =& \frac{1}{\sqrt{30}}(uucc\Bar{c} + \Bar{c}uccu + u\Bar{c}ccu + uu\Bar{c}cc + uuc\Bar{c}c + uccu\Bar{c} + \Bar{c}ccuu + u\Bar{c}cuc + uc\Bar{c}uc + ucc\Bar{c}u \nonumber \\ +& cucu\Bar{c} + \Bar{c}ucuc + c\Bar{c}cuu + cu\Bar{c}uc + cuc\Bar{c}u + cuuc\Bar{c} + \Bar{c}uucc + c\Bar{c}ucu + cu\Bar{c}cu + cuu\Bar{c}c \nonumber \\ +& ccuu\Bar{c} + \Bar{c}cuuc + c\Bar{c}uuc + cc\Bar{c}uu + ccu\Bar{c}u + ucuc\Bar{c} + \Bar{c}cucu + u\Bar{c}ucc + uc\Bar{c}cu +uc\Bar{c}cu + ucu\Bar{c}c)
\end{align}
\end{widetext}
Similarly, we can define this for $uubb\Bar{b}$ pentaquark state. Furthermore, it is feasible to determine the flavor wavefunction for other triply heavy pentaquark states. Moreover, the color wave function associated with a bound state of a hadron is constrained by color confinement, leading to the formation of color singlet states. Additionally, our study focused on $s$- wave triply heavy pentaquark states; therefore, symmetrical restriction for the spatial wave function is trivial. We employed an extended version of the Gursey-Radicati mass formula, along with the effective mass scheme and the screened charge scheme, to determine the masses and magnetic moments of triply heavy pentaquark states. These methodologies are elaborated upon as follows.
\subsection{The Extended Gursey-Radicati mass Formula}
The Gursey-Radicati mass formula proves to be valuable in determining the masses of hadrons. This formula establishes a connection between a particle's mass and its corresponding quantum characteristics, providing valuable insights into the intricate relationship among these foundational attributes. The Gursey-Radicati mass formula was first proposed for the baryon with light quarks ($u$, $d$, $s$) \cite{FG}. It was first extended to include the contribution of charm quarks as \cite{Santo}:
\begin{align}
  M_{GR} = M_0 &+ AS(S+1) + DY  + E[I(I+1) -1/4 Y^2] \nonumber \\ &+ G C_2(SU(3)) + F N_c
\end{align}
Further,  P. Holma et al. modified the  mass formula by including the contribution of the bottom quarks into the counter term as \cite{HOLMA}:
\begin{align}
  M_{GR} = \xi M_0 &+ AS(S+1) + DY  + E[I(I+1)  -1/4 Y^2] \nonumber \\ &+ G C_2(SU(3)) + \sum_{i=c,b} F_i N_i
  \label{mass formula}
  \end{align}

In this context, $M_0$ serves as the scale parameter, while $\xi$ acts as a correction factor to this scale parameter. The value of $\xi$ depends on the constituent quarks (or antiquarks) comprising the hadron. Each quark contributes 1/3 to the total mass of the hadron. Thus, $\xi$ takes on different values: 1 for baryons, 4/3 for tetraquarks, and 5/3 for pentaquarks. $N_i$ represents the counter term for the number of heavy quarks in the hadron. The parameters of the mass formula are extracted from Ref. \cite{Santo, HOLMA} and reported in Table \ref{tab:1}. In the next subsection, we presented the effective mass scheme, which will be used to calculate the masses and magnetic moments of triply heavy pentaquarks.
\begin{table}[ht]
\centering
\caption{Parameters of mass formula with associated uncertainties in MeV \cite{Santo}.}
\tabcolsep 0.4mm  
\begin{tabular}{cccccccc}
       \hline
       \hline
         & $M_0$ & A & D & E & G & $F_c$ & $F_b$ \\
         \hline
        Value & 940.0 & 23.0 & -158.3 & 32.0 & 52.5 & 1354.6  & 4820 \cite{HOLMA} \\
        \hline
        Uncertain.[MeV] & 1.5 & 1.2 & 1.3 & 1.3 & 1.3 & 18.2  & 34.4  \\
        \hline
        \hline
       \end{tabular}
        \label{tab:1}
   \end{table}

\subsection{Effective Mass Scheme}
This study involved a detailed calculation of the effective mass of quarks and anti-quarks, considering their interactions with nearby quarks through a single gluon exchange scheme. The effective mass method enabled us to analyze the complex interaction among quark components, leading to a detailed comprehension of their effective masses in the triply-heavy pentaquark structure. Using these effective quark masses, we performed thorough calculations to establish the masses and magnetic moments of the triply heavy pentaquarks. By incorporating effective mass calculations into our analysis, we obtained a more profound understanding of the complex internal composition of these exotic particles. This study clarifies the mass spectrum of pentaquarks with three heavy quarks and provides insights into their quark dynamics and the strong interactions involved. Therefore, the mass of pentaquarks can be defined as \cite{VERMA}:
\begin{align}
        M_P =& \sum_{i=1}^5 m_i^{ef}  \\
   M_P =& \sum_{i=1}^5 m_i + \sum_{i<j} b_{ij} s_i.s_j
   \label{Effective mass}
\end{align}
 Here, $s_i$ and $s_j$ are the spin operators for the $i^{th}$ and $j^{th}$ quarks (antiquark), and $m_i^{ef}$ represents the effective mass for each quark (antiquark) and $b_{ij}$ are the hyperfine interaction terms. The effective masses equations for constituent quarks inside the pentaquarks are:
 \begin{equation}
 m_1 = m_1 + \alpha b_{12} + \beta b_{13} + \gamma b_{14} + \eta b_{15}
 \end{equation}
 
 \begin{equation}
  m_2 = m_2 + \alpha b_{12} + \beta^{'} b_{23} + \gamma^{'} b_{24} + \eta^{'} b_{25}
  \end{equation}

   \begin{equation}
   m_3 = m_3 + \beta b_{13} + \beta^{'} b_{23} + \gamma^{''} b_{34} + \eta^{''} b_{35} 
   \end{equation}

   \begin{equation}
    m_4 = m_4 + \gamma b_{14} + \gamma^{'} b_{24} + \gamma^{''} b_{34} + \eta^{'''} b_{45} 
 \end{equation}

  \begin{equation}
    m_5 = m_5 + \eta b_{15} + \eta^{'} b_{24} + \eta^{''} b_{34} + \eta^{'''} b_{45} 
 \end{equation}
  Here, 1, 2, 3, 4, and 5 stand for $u$, $d$, $s$, $c$, and $b$ quarks. These equations will get modified if we include two/three/four/five identical quarks \cite{Rohit}. A comprehensive analysis of spin-spin interactions and related factors is written in \cite{sharma2024}. Therefore, effective mass equation for spin-parity $5/2^-$ ($\uparrow\uparrow\uparrow\uparrow\uparrow$) defined as:
\begin{align}
M_{P_{{5/2}^-}} = m_1 + m_2 + m_3 + m_4 + m_5 + \frac{b_{12}}{4}  + \frac{b_{13}}{4}  + \frac{b_{14}}{4} \nonumber \\
 + \frac{b_{15}}{4}  + \frac{b_{23}}{4}  + \frac{b_{24}}{4}  + \frac{b_{25}}{4}  + \frac{b_{34}}{4}  + \frac{b_{35}}{4}  + \frac{b_{45}}{4}
 \label{eff1}
\end{align}
Similarly, for pentaquarks with spin-parity equal to $3/2^-$ ($\uparrow\uparrow\uparrow\uparrow\downarrow$), the effective mass equation can be written as:
\begin{align}
M_{P_{{3/2}^-}} = m_1 + m_2 + m_3 + m_4 + m_5 + \frac{b_{12}}{4}  + \frac{b_{13}}{4}  + \frac{b_{14}}{4} \nonumber \\
 - \frac{b_{15}}{2}  + \frac{b_{23}}{4}  + \frac{b_{24}}{4}  - \frac{b_{25}}{2}  + \frac{b_{34}}{4}  - \frac{b_{35}}{2}  - \frac{b_{45}}{2}
 \label{eff2}
\end{align}
Similarly, for $J^P = 1/2^-$ ($\uparrow\uparrow\uparrow\downarrow\downarrow$), we can write as:
\begin{align}
M_{P_{{1/2}^-}} = m_1 + m_2 + m_3 + m_4 + m_5 + \frac{b_{12}}{4}  + \frac{b_{13}}{4}  - \frac{b_{14}}{2} \nonumber \\
 - \frac{b_{15}}{2}  + \frac{b_{23}}{4}  - \frac{b_{24}}{2}  - \frac{b_{25}}{2}  - \frac{b_{34}}{2}  - \frac{b_{35}}{2}  - \frac{b_{45}}{4}
 \label{eff3}
\end{align}

 Masses of quarks are taken from Ref. \cite{Rohit}. Hyperfine interaction terms are calculated using the effective mass equations \eqref{eff1}, \eqref{eff2}, and \eqref{eff3} by taking the masses obtained from the Gursey-Radicati mass formula as scale and written as:
\begin{align}
     m_u = m_d = \hspace{0.3cm} 362 MeV, \hspace{0.3cm} m_s = \hspace{0.3cm} 539 MeV \nonumber \\
    m_c = \hspace{0.3cm} 1710 MeV, \hspace{0.3cm} m_b = \hspace{0.3cm} 5043 MeV
\end{align}
\begin{equation}
     b_{uu} = \hspace{0.3cm} b_{ud} = \hspace{0.3cm} b_{dd} = 119.70 MeV
 \end{equation}

 \begin{equation}
     b_{us} = \hspace{0.3cm} b_{ds} \hspace{0.3cm} = \left(\frac{m_u}{m_s}\right)b_{uu} = 80.12 MeV
 \end{equation}

\begin{equation}
     b_{uc} = \hspace{0.3cm} b_{dc} = \hspace{0.3cm} \left(\frac{m_u}{m_c}\right)b_{uu} = 25.20 MeV
 \end{equation}

\begin{equation}
     b_{ss} = \hspace{0.3cm} \left(\frac{m_u}{m_s}\right)^2 b_{uu} =  53.80 MeV
 \end{equation}

\begin{equation}
     b_{sc} = \hspace{0.3cm} \left(\frac{m_u^2}{m_s m_c}\right) b_{uu} = 16.90 MeV
\end{equation}

\begin{equation}
     b_{cc} = \hspace{0.3cm} \left(\frac{m_u}{ m_c}\right)^2 b_{uu} = 5.33 MeV
 \end{equation}

\begin{equation}
     b_{ub} = \hspace{0.3cm} b_{db} = \hspace{0.3cm} \left(\frac{m_u}{m_b}\right)b_{uu}  = 8.59 MeV
 \end{equation}

\begin{equation}
     b_{bb} = \hspace{0.3cm} \left(\frac{m_u}{m_b}\right)^2 b_{uu} =  0.61 MeV
 \end{equation}

 \begin{equation}
     b_{sb} = \hspace{0.3cm} \left(\frac{m_u}{m_s}\right) b_{ub} =  5.76 MeV
 \end{equation}

 \begin{equation}
     b_{cb} = \hspace{0.3cm} \left(\frac{m_u}{m_c}\right) b_{ub} =  1.81 MeV
 \end{equation}
 by using these values of hyperfine interaction terms in the total effective mass equations \eqref{eff1}, \eqref{eff2} and \eqref{eff3}, we calculated the masses and magnetic moments of triply heavy pentaquarks. Effective quark masses are reported in Tables \ref{tab:em1} and \ref{tab:em2}. The masses of triply heavy pentaquarks are written in Tables \ref{tab:mass1} and \ref{tab:mass3}, whereas magnetic moment assignments are written in Table \ref{tab: magnetic moment1} and \ref{tab: magnetic moments}. In the next section, we introduced the screened charge scheme for triply heavy pentaquarks. 

\begin{table*}\renewcommand{\arraystretch}{0.8}
 \tabcolsep 0.5mm
\centering
\caption{Table of effective quark masses for $q\overline{q}ccc$, $q\overline{q}bbb$, $qqcc\overline{c}$, and $qqbb\overline{b}$ - type pentaquarks with various $J^P$ values. The effective quark masses for up, down, strange, charm, anti-charm, and bottom quarks are $m_u*$, $m_d*$, $m_s*$, $m_c*$, and $m_b*$, respectively, in MeV.}
\begin{tabular}{ccccccccc}
       \hline
       \hline
         & \hspace{0.3cm} $m_u^*$/$m_d^*$ & \hspace{0.3cm} $\Bar{m_u^*}$/$\Bar{m_d^*}$ & \hspace{0.3cm}$m_s^*$ & \hspace{0.3cm}$\Bar{m_s^*}$ & \hspace{0.3cm}$m_c^*$ & \hspace{0.3cm}$m_{\Bar{c}}^*$ & \hspace{0.3cm}$m_b^*$&\hspace{0.3cm}$m_{\Bar{b}}^*$ \\
         \hline
\underline{$q\overline{q}ccc$-type Pentaquarks with $J^P = 1/2^-$} &  &   &  &  &\\
      $s=0$ & \hspace{0.3cm} 328.13  & \hspace{0.3cm} 328.13 & \hspace{0.3cm} - & \hspace{0.3cm} - & \hspace{0.3cm} 1698.73 & \hspace{0.3cm} - & \hspace{0.3cm} - & -\\
      $s=1$ & \hspace{0.3cm} 333.08 & \hspace{0.3cm} - & \hspace{0.3cm} - & \hspace{0.3cm} 516.31 & \hspace{0.3cm} 1700.81 & \hspace{0.3cm} - & \hspace{0.3cm} - & -\\
      $s=2$ & \hspace{0.3cm} - & \hspace{0.3cm} - & \hspace{0.3cm} 519.60 & \hspace{0.3cm} 519.60  & \hspace{0.3cm} 1702.88 & \hspace{0.3cm} - & \hspace{0.3cm} - & - \\
\underline{$q\overline{q}ccc$-type Pentaquarks with $J^P = 3/2^-$} &  &   &  &  &\\
      $s=0$ & \hspace{0.3cm} 341.52  & \hspace{0.3cm} 341.52 & \hspace{0.3cm} - & \hspace{0.3cm} - & \hspace{0.3cm} 1708.19 & \hspace{0.3cm} - & \hspace{0.3cm} - & -\\
      $s=1$ & \hspace{0.3cm} 351.42 & \hspace{0.3cm} - & \hspace{0.3cm} - & \hspace{0.3cm} 506.29 & \hspace{0.3cm} 1710.26 & \hspace{0.3cm} - & \hspace{0.3cm} - & -\\
      $s=2$ & \hspace{0.3cm} - & \hspace{0.3cm} - & \hspace{0.3cm} 531.88 & \hspace{0.3cm} 531.88  & \hspace{0.3cm} 1709.22 & \hspace{0.3cm} - & \hspace{0.3cm} - & - \\
 \underline{$q\overline{q}ccc$-type Pentaquarks with $J^P = 5/2^-$} &  &   &  &  &\\
      $s=0$ & \hspace{0.3cm} 386.41  & \hspace{0.3cm} 386.41 & \hspace{0.3cm} - & \hspace{0.3cm} - & \hspace{0.3cm} 1717.63 & \hspace{0.3cm} - & \hspace{0.3cm} - & -\\
      $s=1$ & \hspace{0.3cm} 381.46 & \hspace{0.3cm} - & \hspace{0.3cm} - & \hspace{0.3cm} 555.35 & \hspace{0.3cm} 1716.60 & \hspace{0.3cm} - & \hspace{0.3cm} - & -\\
      $s=2$ & \hspace{0.3cm} - & \hspace{0.3cm} - & \hspace{0.3cm} 552.06 & \hspace{0.3cm} 552.06  & \hspace{0.3cm} 1715.56 & \hspace{0.3cm} - & \hspace{0.3cm} - & - \\
\hline
\underline{$q\overline{q}bbb$-type Pentaquarks with $J^P = 1/2^-$} &  &   &  & & \\
  $s=0$ & \hspace{0.3cm} 340.59  & \hspace{0.3cm} 340.59 & \hspace{0.3cm} - & \hspace{0.3cm} - & \hspace{0.3cm} - & \hspace{0.3cm} - & \hspace{0.3cm} 5038.86 & \hspace{0.3cm} - \\
  $s=1$ & \hspace{0.3cm} 345.54 & \hspace{0.3cm} - & \hspace{0.3cm} - &  \hspace{0.3cm} 524.66 & \hspace{0.3cm} - & \hspace{0.3cm} - & \hspace{0.3cm} 5039.57 & \hspace{0.3cm} -  \\
  $s=2$ & \hspace{0.3cm} - & \hspace{0.3cm} - & \hspace{0.3cm} 527.95 & \hspace{0.3cm} 527.95 & \hspace{0.3cm}  & \hspace{0.3cm} - & \hspace{0.3cm} 5040.27 & \hspace{0.3cm} -  \\
\underline{$q\overline{q}bbb$-type Pentaquarks with $J^P = 3/2^-$} &  &   &  & & \\
     $s=0$ & \hspace{0.3cm} 363.07  & \hspace{0.3cm} 363.07 & \hspace{0.3cm} - & \hspace{0.3cm} - & \hspace{0.3cm} - & \hspace{0.3cm} - & \hspace{0.3cm} 5042.08 & \hspace{0.3cm} - \\
      $s=1$ & \hspace{0.3cm} 345.19 & \hspace{0.3cm} - & \hspace{0.3cm} - &  \hspace{0.3cm} 544.65 & \hspace{0.3cm} - & \hspace{0.3cm} - & \hspace{0.3cm} 5042.79 & \hspace{0.3cm} -  \\
      $s=2$ & \hspace{0.3cm} - & \hspace{0.3cm} - & \hspace{0.3cm} 527.71 & \hspace{0.3cm} 527.71 & \hspace{0.3cm}  & \hspace{0.3cm} - & \hspace{0.3cm} 5042.43 & \hspace{0.3cm} -  \\
\underline{$q\overline{q}bbb$-type Pentaquarks with $J^P = 5/2^-$} &  &   &  & & \\
     $s=0$ & \hspace{0.3cm} 380.18 & \hspace{0.3cm} 380.18 & \hspace{0.3cm} - & \hspace{0.3cm} - & \hspace{0.3cm} - & \hspace{0.3cm} - & \hspace{0.3cm} 5045.30 & \hspace{0.3cm} - \\
      $s=1$ & \hspace{0.3cm} 375.23 & \hspace{0.3cm} - & \hspace{0.3cm} - &  \hspace{0.3cm} 551.17 & \hspace{0.3cm} - & \hspace{0.3cm} - & \hspace{0.3cm} 5044.95 & \hspace{0.3cm} -  \\
      $s=2$ & \hspace{0.3cm} - & \hspace{0.3cm} - & \hspace{0.3cm} 547.88 & \hspace{0.3cm} 547.88 & \hspace{0.3cm}  & \hspace{0.3cm} - & \hspace{0.3cm} 5044.59 & -  \\
      \hline
      \hline
\underline{$qqcc\overline{c}$-type Pentaquarks with $J^P = 1/2^-$} &  &   &  &  &\\
      $s=0$ & \hspace{0.3cm} 367.51  & \hspace{0.3cm} - & \hspace{0.3cm} - & \hspace{0.3cm} - & \hspace{0.3cm} 1713.64 & \hspace{0.3cm} 1713.64 & \hspace{0.3cm} - & -\\
      $s=1$ & \hspace{0.3cm} 362.56 & \hspace{0.3cm} - & \hspace{0.3cm} 542.67 & \hspace{0.3cm} - & \hspace{0.3cm} 1712.60 & \hspace{0.3cm} 1712.64 & \hspace{0.3cm} - & -\\
      $s=2$ & \hspace{0.3cm} - & \hspace{0.3cm} - & \hspace{0.3cm} 539.38 & \hspace{0.3cm} -  & \hspace{0.3cm} 1711.56 & \hspace{0.3cm} 1711.56 & \hspace{0.3cm} - & - \\
\underline{$qqcc\overline{c}$-type Pentaquarks with $J^P = 3/2^-$} &  &   &  &  &\\
      $s=0$ & \hspace{0.3cm} 376.96  & \hspace{0.3cm} - & \hspace{0.3cm} - & \hspace{0.3cm} - & \hspace{0.3cm} 1715.63 & \hspace{0.3cm} 1715.63 & \hspace{0.3cm} - & -\\
      $s=1$ & \hspace{0.3cm} 372.01 & \hspace{0.3cm} - & \hspace{0.3cm} 549.01 & \hspace{0.3cm} - & \hspace{0.3cm} 1714.60 & \hspace{0.3cm} 1714.60 & \hspace{0.3cm} - & -\\
      $s=2$ & \hspace{0.3cm} - & \hspace{0.3cm} - & \hspace{0.3cm} 545.72 & \hspace{0.3cm} - & \hspace{0.3cm} 1698.89 & \hspace{0.3cm} 1698.89 & \hspace{0.3cm} - & - \\
\underline{$qqcc\overline{c}$-type Pentaquarks with $J^P = 5/2^-$} &  &   &  &  &\\
      $s=0$ & \hspace{0.3cm} 386.41  & \hspace{0.3cm} - & \hspace{0.3cm} - & \hspace{0.3cm} - & \hspace{0.3cm} 1717.63 & \hspace{0.3cm} 1717.63 & \hspace{0.3cm} - & -\\
      $s=1$ & \hspace{0.3cm} 381.46 & \hspace{0.3cm} - & \hspace{0.3cm} 555.35 & \hspace{0.3cm} - & \hspace{0.3cm} 1716.60 & \hspace{0.3cm} 1716.60 & \hspace{0.3cm} - & -\\
      $s=2$ & \hspace{0.3cm} - & \hspace{0.3cm} - & \hspace{0.3cm} 552.06 & \hspace{0.3cm} -  & \hspace{0.3cm} 1715.56 & \hspace{0.3cm} 1715.56 & \hspace{0.3cm} - & - \\
  \hline
\underline{$qqbb\overline{b}$-type Pentaquarks with $J^P = 1/2^-$} &  &   &  & & \\
  $s=0$ & \hspace{0.3cm} 340.59  & \hspace{0.3cm} 340.59 & \hspace{0.3cm} - & \hspace{0.3cm} - & \hspace{0.3cm} - & \hspace{0.3cm} - & \hspace{0.3cm} 5038.86 & \hspace{0.3cm} - \\
  $s=1$ & \hspace{0.3cm} 345.54 & \hspace{0.3cm} - & \hspace{0.3cm} - &  \hspace{0.3cm} 524.66 & \hspace{0.3cm} - & \hspace{0.3cm} - & \hspace{0.3cm} 5039.57 & \hspace{0.3cm} -  \\
  $s=2$ & \hspace{0.3cm} - & \hspace{0.3cm} - & \hspace{0.3cm} 527.95 & \hspace{0.3cm} 527.95 & \hspace{0.3cm}  & \hspace{0.3cm} - & \hspace{0.3cm} 5040.27 & \hspace{0.3cm} -  \\
\underline{$qqbb\overline{b}$-type Pentaquarks with $J^P = 3/2^-$} &  &   &  & & \\
     $s=0$ & \hspace{0.3cm} 363.07  & \hspace{0.3cm} 363.07 & \hspace{0.3cm} - & \hspace{0.3cm} - & \hspace{0.3cm} - & \hspace{0.3cm} - & \hspace{0.3cm} 5042.08 & \hspace{0.3cm} - \\
      $s=1$ & \hspace{0.3cm} 345.19 & \hspace{0.3cm} - & \hspace{0.3cm} - &  \hspace{0.3cm} 544.65 & \hspace{0.3cm} - & \hspace{0.3cm} - & \hspace{0.3cm} 5042.79 & \hspace{0.3cm} -  \\
      $s=2$ & \hspace{0.3cm} - & \hspace{0.3cm} - & \hspace{0.3cm} 527.71 & \hspace{0.3cm} 527.71 & \hspace{0.3cm}  & \hspace{0.3cm} - & \hspace{0.3cm} 5042.43 & \hspace{0.3cm} -  \\
\underline{$qqbb\overline{b}$-type Pentaquarks with $J^P = 5/2^-$} &  &   &  & & \\
     $s=0$ & \hspace{0.3cm} 380.18 & \hspace{0.3cm} - & \hspace{0.3cm} - & \hspace{0.3cm} - & \hspace{0.3cm} - & \hspace{0.3cm} - & \hspace{0.3cm} 5045.30 & \hspace{0.3cm} 5045.30 \\
      $s=1$ & \hspace{0.3cm} 375.23 & \hspace{0.3cm} - & \hspace{0.3cm} 551.75 &  \hspace{0.3cm} - & \hspace{0.3cm} - & \hspace{0.3cm} - & \hspace{0.3cm} 5044.95 & \hspace{0.3cm} 5044.95  \\
      $s=2$ & \hspace{0.3cm} - & \hspace{0.3cm} - & \hspace{0.3cm} 547.88 & \hspace{0.3cm} 547.88 & \hspace{0.3cm}  & \hspace{0.3cm} - & \hspace{0.3cm} 5044.59 & \hspace{0.3cm} 5044.59 \\
      \hline
      \hline
       \end{tabular}
        \label{tab:em1}
   \end{table*}

\begin{table*}\renewcommand{\arraystretch}{0.8}
 \tabcolsep 0.5mm
\centering
\caption{Table for the effective quark masses for $q\overline{q}ccb$, $q\overline{q}bbc$, $qqcc\overline{b}$, and $qqbb\overline{c}$ - type pentaquarks with different possible $J^P$ values. Here, $m_u^*$, $m_d^*$, $m_s^*$ $m_c^*$, $m_{\Bar{c}}^*$ and $m_b^*$ are the effective quark masses for respective quarks. All masses are in the units of MeV.}
\begin{tabular}{ccccccccc}
       \hline
       \hline
         & \hspace{0.3cm} $m_u^*$/$m_d^*$ & \hspace{0.3cm} $\Bar{m_u^*}$/$\Bar{m_d^*}$ & \hspace{0.3cm}$m_s^*$ & \hspace{0.3cm}$\Bar{m_s^*}$ & \hspace{0.3cm}$m_c^*$ & \hspace{0.3cm}$m_{\Bar{c}}^*$ & \hspace{0.3cm}$m_b^*$&\hspace{0.3cm}$m_{\Bar{b}}^*$ \\
         \hline
\underline{$q\overline{q}ccb$-type Pentaquarks with $J^P = 1/2^-$} &  &   &  &  &\\
      $s=0$ & \hspace{0.3cm} 332.29  & \hspace{0.3cm} 332.29 & \hspace{0.3cm} - & \hspace{0.3cm} - & \hspace{0.3cm} 1698.29 & \hspace{0.3cm} - & \hspace{0.3cm} - & \hspace{0.3cm} 5039.16\\
      $s=1$ & \hspace{0.3cm} 337.23 & \hspace{0.3cm} - & \hspace{0.3cm} - & \hspace{0.3cm} 519.09 & \hspace{0.3cm} 1700.37 & \hspace{0.3cm} - & \hspace{0.3cm} - & \hspace{0.3cm} 5039.87\\
      $s=2$ & \hspace{0.3cm} - & \hspace{0.3cm} - & \hspace{0.3cm} 522.38 & \hspace{0.3cm} 522.38  & \hspace{0.3cm} 1702.44 & \hspace{0.3cm} - & \hspace{0.3cm} - & \hspace{0.3cm} 5040.57 \\
\underline{$q\overline{q}ccb$-type Pentaquarks with $J^P = 3/2^-$} &  &   &  &  &\\
      $s=0$ & \hspace{0.3cm} 339.44  & \hspace{0.3cm} 339.41 & \hspace{0.3cm} - & \hspace{0.3cm} - & \hspace{0.3cm} 1707.74 & \hspace{0.3cm} - & \hspace{0.3cm} - & \hspace{0.3cm} 5042.38\\
      $s=1$ & \hspace{0.3cm} 349.44 & \hspace{0.3cm} - & \hspace{0.3cm} - & \hspace{0.3cm} 509.08 & \hspace{0.3cm} 1709.82 & \hspace{0.3cm} - & \hspace{0.3cm} - & \hspace{0.3cm} 5043.09\\
      $s=2$ & \hspace{0.3cm} - & \hspace{0.3cm} - & \hspace{0.3cm} 530.49 & \hspace{0.3cm} 530.49  & \hspace{0.3cm} 1708.78 & \hspace{0.3cm} - & \hspace{0.3cm} - & \hspace{0.3cm} 5042.73 \\
 \underline{$q\overline{q}ccb$-type Pentaquarks with $J^P = 5/2^-$} &  &   &  &  &\\
      $s=0$ & \hspace{0.3cm} 384.33  & \hspace{0.3cm} 384.33 & \hspace{0.3cm} - & \hspace{0.3cm} - & \hspace{0.3cm} 1717.19 & \hspace{0.3cm} - & \hspace{0.3cm} - & \hspace{0.3cm} 5045.60\\
      $s=1$ & \hspace{0.3cm} 379.89 & \hspace{0.3cm} - & \hspace{0.3cm} - & \hspace{0.3cm} 553.96 & \hspace{0.3cm} 1716.16 & \hspace{0.3cm} - & \hspace{0.3cm} - & \hspace{0.3cm} 5045.25\\
      $s=2$ & \hspace{0.3cm} - & \hspace{0.3cm} - & \hspace{0.3cm} 550.67 & \hspace{0.3cm} 550.67  & \hspace{0.3cm} 1715.20 & \hspace{0.3cm} - & \hspace{0.3cm} - & \hspace{0.3cm} 5044.89 \\
\hline
\underline{$q\overline{q}bbc$-type Pentaquarks with $J^P = 1/2^-$} &  &   &  & & \\
  $s=0$ & \hspace{0.3cm} 336.44  & \hspace{0.3cm} 336.44 & \hspace{0.3cm} - & \hspace{0.3cm} - & \hspace{0.3cm} - & \hspace{0.3cm} 1697.85 & \hspace{0.3cm} 5039.01 & \hspace{0.3cm} - \\
  $s=1$ & \hspace{0.3cm} 341.39 & \hspace{0.3cm} - & \hspace{0.3cm} - &  \hspace{0.3cm} 521.88 & \hspace{0.3cm} - & \hspace{0.3cm} 1699.93 & \hspace{0.3cm} 5039.72 & \hspace{0.3cm} -  \\
  $s=2$ & \hspace{0.3cm} - & \hspace{0.3cm} - & \hspace{0.3cm} 525.17 & \hspace{0.3cm} 525.17 & \hspace{0.3cm} - & \hspace{0.3cm} 1702.00 & \hspace{0.3cm} 5040.42 & \hspace{0.3cm} -  \\
\underline{$q\overline{q}bbc$-type Pentaquarks with $J^P = 3/2^-$} &  &   &  & & \\
     $s=0$ & \hspace{0.3cm} 337.37  & \hspace{0.3cm} 337.37 & \hspace{0.3cm} - & \hspace{0.3cm} - & \hspace{0.3cm} - & \hspace{0.3cm} 1707.30 & \hspace{0.3cm} 5043.23 & \hspace{0.3cm} - \\
      $s=1$ & \hspace{0.3cm} 347.26 & \hspace{0.3cm} - & \hspace{0.3cm} - &  \hspace{0.3cm} 511.86 & \hspace{0.3cm} - & \hspace{0.3cm} 1709.38 & \hspace{0.3cm} 5042.94 & \hspace{0.3cm} -  \\
      $s=2$ & \hspace{0.3cm} - & \hspace{0.3cm} - & \hspace{0.3cm} 529.10 & \hspace{0.3cm} 529.10 & \hspace{0.3cm}  & \hspace{0.3cm} 1708.34 & \hspace{0.3cm} 5042.58 & \hspace{0.3cm} -  \\
\underline{$q\overline{q}bbc$-type Pentaquarks with $J^P = 5/2^-$} &  &   &  & & \\
     $s=0$ & \hspace{0.3cm} 382.26 & \hspace{0.3cm} 382.26 & \hspace{0.3cm} - & \hspace{0.3cm} - & \hspace{0.3cm} - & \hspace{0.3cm} 1716.75 & \hspace{0.3cm} 5045.45 & \hspace{0.3cm} - \\
      $s=1$ & \hspace{0.3cm} 377.31 & \hspace{0.3cm} - & \hspace{0.3cm} - &  \hspace{0.3cm} 552.56 & \hspace{0.3cm} - & \hspace{0.3cm} 1715.72 & \hspace{0.3cm} 5045.10 & \hspace{0.3cm} -  \\
      $s=2$ & \hspace{0.3cm} - & \hspace{0.3cm} - & \hspace{0.3cm} 549.27 & \hspace{0.3cm} 549.27 & \hspace{0.3cm} - & \hspace{0.3cm} 1714.68 & \hspace{0.3cm} 5044.70 & \hspace{0.3cm} - \\
      \hline
      \hline
\underline{$qqcc\overline{b}$-type Pentaquarks with $J^P = 1/2^-$} &  &   &  &  &\\
      $s=0$ & \hspace{0.3cm} 371.66  & \hspace{0.3cm} - & \hspace{0.3cm} - & \hspace{0.3cm} - & \hspace{0.3cm} 1714.52 & \hspace{0.3cm} - & \hspace{0.3cm} - & \hspace{0.3cm} 5038.03\\
      $s=1$ & \hspace{0.3cm} 366.71 & \hspace{0.3cm} - & \hspace{0.3cm} 541.75 & \hspace{0.3cm} - & \hspace{0.3cm} 1713.48 & \hspace{0.3cm} - & \hspace{0.3cm} - & \hspace{0.3cm} 5038.73\\
      $s=2$ & \hspace{0.3cm} - & \hspace{0.3cm} - & \hspace{0.3cm} 542.17 & \hspace{0.3cm} -  & \hspace{0.3cm} 1712.44 & \hspace{0.3cm} - & \hspace{0.3cm} - & \hspace{0.3cm} 5039.44\\
\underline{$qqcc\overline{b}$-type Pentaquarks with $J^P = 3/2^-$} &  &   &  &  &\\
      $s=0$ & \hspace{0.3cm} 381.11  & \hspace{0.3cm} - & \hspace{0.3cm} - & \hspace{0.3cm} - & \hspace{0.3cm} 1716.51 & \hspace{0.3cm} - & \hspace{0.3cm} - & \hspace{0.3cm} 5037.80\\
      $s=1$ & \hspace{0.3cm} 376.16 & \hspace{0.3cm} - & \hspace{0.3cm} 551.80 & \hspace{0.3cm} - & \hspace{0.3cm} 1715.48 & \hspace{0.3cm} - & \hspace{0.3cm} - & \hspace{0.3cm} 5038.51\\
      $s=2$ & \hspace{0.3cm} - & \hspace{0.3cm} - & \hspace{0.3cm} 548.51 & \hspace{0.3cm} - & \hspace{0.3cm} 1714.44 & \hspace{0.3cm} - & \hspace{0.3cm} - & \hspace{0.3cm} 5039.22 \\
\underline{$qqcc\overline{b}$-type Pentaquarks with $J^P = 5/2^-$} &  &   &  &  &\\
      $s=0$ & \hspace{0.3cm} 384.33  & \hspace{0.3cm} - & \hspace{0.3cm} - & \hspace{0.3cm} - & \hspace{0.3cm} 1717.19 & \hspace{0.3cm} - & \hspace{0.3cm} - & \hspace{0.3cm} 5045.60\\
      $s=1$ & \hspace{0.3cm} 379.38 & \hspace{0.3cm} - & \hspace{0.3cm} 553.96 & \hspace{0.3cm} - & \hspace{0.3cm} 1716.16 & \hspace{0.3cm} - & \hspace{0.3cm} - & \hspace{0.3cm} 5045.25\\
      $s=2$ & \hspace{0.3cm} - & \hspace{0.3cm} - & \hspace{0.3cm} 550.67 & \hspace{0.3cm} -  & \hspace{0.3cm} 1715.12 & \hspace{0.3cm} - & \hspace{0.3cm} - & \hspace{0.3cm} 5044.89 \\
  \hline
\underline{$qqbb\overline{c}$-type Pentaquarks with $J^P = 1/2^-$} &  &   &  & & \\
  $s=0$ & \hspace{0.3cm} 369.58  & \hspace{0.3cm} - & \hspace{0.3cm} - & \hspace{0.3cm} - & \hspace{0.3cm} - & \hspace{0.3cm} 1696.72 & \hspace{0.3cm} 5044.54 & \hspace{0.3cm} - \\
  $s=1$ & \hspace{0.3cm} 364.64 & \hspace{0.3cm} - & \hspace{0.3cm} 544.07 &  \hspace{0.3cm} - & \hspace{0.3cm} - & \hspace{0.3cm} 1698.80 & \hspace{0.3cm} 5044.19 & \hspace{0.3cm} -  \\
  $s=2$ & \hspace{0.3cm} - & \hspace{0.3cm} - & \hspace{0.3cm} 540.78 & \hspace{0.3cm} - & \hspace{0.3cm} - & \hspace{0.3cm} 1700.87 & \hspace{0.3cm} 5043.83 & \hspace{0.3cm} -  \\
\underline{$qqbb\overline{c}$-type Pentaquarks with $J^P = 3/2^-$} &  &   &  & & \\
     $s=0$ & \hspace{0.3cm} 372.81  & \hspace{0.3cm} - & \hspace{0.3cm} - & \hspace{0.3cm} - & \hspace{0.3cm} - & \hspace{0.3cm} 1696.50 & \hspace{0.3cm} 5044.77 & \hspace{0.3cm} - \\
      $s=1$ & \hspace{0.3cm} 367.86 & \hspace{0.3cm} - & \hspace{0.3cm} 546.23 &  \hspace{0.3cm} - & \hspace{0.3cm} - & \hspace{0.3cm} 1698.57 & \hspace{0.3cm} 5044.42 & \hspace{0.3cm} -  \\
      $s=2$ & \hspace{0.3cm} - & \hspace{0.3cm} - & \hspace{0.3cm} 542.94 & \hspace{0.3cm} - & \hspace{0.3cm} - & \hspace{0.3cm} 1700.65 & \hspace{0.3cm} 5044.06 & \hspace{0.3cm} -  \\
\underline{$qqbb\overline{c}$-type Pentaquarks with $J^P = 5/2^-$} &  &   &  & & \\
     $s=0$ & \hspace{0.3cm} 382.26 & \hspace{0.3cm} - & \hspace{0.3cm} - & \hspace{0.3cm} - & \hspace{0.3cm} - & \hspace{0.3cm} 1716.75 & \hspace{0.3cm} 5045.45 & \hspace{0.3cm} - \\
      $s=1$ & \hspace{0.3cm} 377.31 & \hspace{0.3cm} - & \hspace{0.3cm} 552.56 &  \hspace{0.3cm} - & \hspace{0.3cm} - & \hspace{0.3cm} 1715.72 & \hspace{0.3cm} 5045.10 & \hspace{0.3cm} - \\
      $s=2$ & \hspace{0.3cm} - & \hspace{0.3cm} - & \hspace{0.3cm} 549.27 & \hspace{0.3cm} - & \hspace{0.3cm} - & \hspace{0.3cm} 1714.68 & \hspace{0.3cm} 5044.74 & \hspace{0.3cm} - \\
      \hline
      \hline
       \end{tabular}
        \label{tab:em2}
   \end{table*}

 \begin{table*}[ht!]
        \centering
         \caption{Table for masses of triply heavy $q\overline{q}ccc$ or $q\overline{q}bbb$ and $qqcc\Bar{c}$ or $qqbb\Bar{b}$ - type pentaquarks in octet and sextet configuration of SU(3) flavor representation by using the formalism of GR mass formula and the effective mass scheme. All masses are in MeV.}
          \begin{tabular}{cccccccccc}
      \hline
      \hline
       \textcolor{red}{$q\overline{q}QQQ$ - type Pentaquarks} & & &  &  &  & & \\
      \hline
   Quark Content &  S &  I &  Y & $SU(3)_f$ & $C_2(SU(3)_f)$  & GR formula & Eff. mass Scheme & Ref. \cite{EPJC2} & Ref. \cite{Reference}\\
   \hline
$u\overline{u}ccc$, $u\overline{d}ccc$, $d\overline{d}ccc$ & 1/2 & 0 & 0 & $[21]_8$ & 3 &  5805.15 $\pm$ 54.76 & 5752.47 & 5910 & 5741\\ 
    & 1/2 & 1 & 0 & $[21]_8$ & 3 &  5869.15 $\pm$ 54.82 &  &  6047 & 5963\\ 
   & 3/2 & 0 & 0 & $[21]_8$ &  3 & 5874.15 $\pm$ 54.94 & 5779.25 & & 5786\\
   & 3/2 & 1 & 0 & $[21]_8$ &  3 & 5938.15 $\pm$ 55.00 &  & 5922 & 5854\\
  & 5/2 & 0 & 0 & $[21]_8$ & 3  & 5989.15 $\pm$ 55.75 & 5925.72 &  & 5786\\
   & 5/2 & 1 & 0 & $[21]_8$ & 3  & 6053.15 $\pm$ 55.81 & & \\
     \hline
     $u\overline{s}ccc$, $d \overline{s}ccc$ & 1/2 & 1/2 & 1 & $[21]_8$ & 3 & 5662.85 $\pm$ 54.78  & 5951.82 & 6086 &\\ 
     & 3/2 & 1/2 & 1 & $[21]_8$ &  3 & 5731.85 $\pm$ 54.96 & 5988.49 & 6107\\
      & 5/2 & 1/2  & 1 & $[21]_8$ &  3 & 5846.85 $\pm$ 55.76 & 6086.60 & \\
     \hline
 $s\overline{u}ccc$, $s \overline{d}ccc$ & 1/2 & 1/2 & -1 & $[21]_8$ & 3 & 5979.45 $\pm$ 54.78  & 5951.82 & 6082\\ 
     & 3/2 & 1/2 & -1 & $[21]_8$ &  3 & 6048.45 $\pm$ 54.96 & 5988.49 & 6109\\
      & 5/2 & 1/2  & -1 & $[21]_8$ &  3 & 6163.45 $\pm$ 55.77 & 6086.60 & \\
    \hline
    $s\overline{s}ccc$ & 1/2 & 1/2 & 0 & $[21]_8$ & 3 & 5805.15 $\pm$ 54.76 & 6147.85  & 6254\\ 
     & 3/2 & 1/2 & 0 & $[21]_8$ &  3 & 5874.15 $\pm$ 54.94 & 6172.42 & 6296\\
      & 5/2 & 1/2  & 0 & $[21]_8$ &  3 & 5989.15 $\pm$ 55.75 & 6250.80 & \\
     \hline
    $u\overline{u}bbb$, $u \overline{d}bbb$, $d \overline{d}bbb$  & 1/2 & 0 & 0 & $[21]_8$ & 3 & 16201.40 $\pm$ 103.28 & 15797.80 & 15891 & 16032\\ 
    & 1/2 & 1 & 0 & $[21]_8$ & 3 &  16265.40 $\pm$ 103.32 & & 15975 & 15957\\ 
   & 3/2 & 0 & 0 & $[21]_8$ &  3 & 16270.40 $\pm$ 103.38 & 15842.70 & 15888 & 15796\\
   & 3/2 & 1 & 0 & $[21]_8$ &  3 & 16334.40 $\pm$ 103.41 &  &  & 15967\\
  & 5/2 & 0 & 0 & $[21]_8$ & 3  & 16385.40 $\pm$ 103.81 & 15900.40 &  & 15796\\
   & 5/2 & 1 & 0 & $[21]_8$ & 3  & 16449.40 $\pm$ 103.84 &  & \\
   \hline
     $u \overline{s}bbb$, $d \overline{s}bbb$ & 1/2 & 1/2 & 1 & $[21]_8$ & 3 & 16059.10 $\pm$ 103.29  & 15988.90 & 16135\\ 
     & 3/2 & 1/2 & 1 & $[21]_8$ &  3 & 16128.10 $\pm$ 103.39 & 15988.20 & 16074\\
      & 5/2 & 1/2  & 1 & $[21]_8$ &  3 & 16243.10 $\pm$ 103.82 & 16061.30 & \\
     \hline
$s\overline{u}bbb$, $s \overline{d}bbb$  & 1/2 & 1/2 & -1 & $[21]_8$ & 3 & 16375.70 $\pm$ 103.29  & 15988.90 & 16134\\ 
     & 3/2 & 1/2 & -1 & $[21]_8$ &  3 & 16444.70 $\pm$ 103.39 & 15988.20 & 16074\\
      & 5/2 & 1/2  & -1 & $[21]_8$ &  3 & 16559.70 $\pm$ 103.82 &  16061.30 & \\
     \hline
     $s\overline{s}bbb$ & 1/2 & 1/2 & 0 & $[21]_8$ & 3 & 16201.40 $\pm$ 103.28  & 16176.70 & 16255\\ 
     & 3/2 & 1/2 & 0 & $[21]_8$ &  3 & 16270.40 $\pm$ 103.38 & 16176.20 & 16260\\
      & 5/2 & 1/2  & 0 & $[21]_8$ &  3 & 16385.40 $\pm$ 103.81 & 16229.60 & \\
     \hline
       \textcolor{red}{$qqQQ\overline{Q}$ - type Pentaquarks} &  &  &  &  &  &\\
     \hline
    $uucc\overline{c}$, $udcc\overline{c}$, $ddcc\overline{c}$  & 1/2 & 0 & 2/3 & $[2]_6$ & 10/3 &  5713.56 $\pm$ 55.79 & \\ 
    & 1/2 & 1 & 2/3 & $[2]_6$ & 10/3 &  5777.56 $\pm$ 54.86 & 5839.46\\ 
   & 3/2 & 0 & 2/3 & $[2]_6$ &  10/3 & 5782.56 $\pm$ 54.98 &  & \\
   & 3/2 & 1 & 2/3 & $[2]_6$ &  10/3 & 5846.56 $\pm$ 55.03 & 5879.93 & \\
  & 5/2 & 0 & 2/3 & $[2]_6$ & 10/3  & 5897.56 $\pm$ 55.79 &  & \\
   & 5/2 & 1 & 2/3 & $[2]_6$ & 10/3  & 5961.56 $\pm$ 55.85 & 5925.72 & \\
     \hline
$uscc\overline{c}$, $dscc\overline{c}$  & 1/2 & 1/2 & -1/3 & $[2]_6$ & 10/3 & 5898.53 $\pm$ 54.80  & 6012.79 & \\ 
& 3/2 & 1/2 & -1/3 & $[2]_6$ &  10/3 & 5967.53 $\pm$ 54.98 & 6047.03 & \\
  & 5/2 & 1/2  & -1/3 & $[2]_6$ &  10/3 & 6082.53 $\pm$ 55.79 & 6086.60 & \\
     \hline
    $sscc\overline{c}$ & 1/2  & 0  & -4/3 & $[2]_6$ & 10/3 & 6019.49 $\pm$ 54.83 & 6189.44 & \\ 
     & 3/2 & 0 & -4/3 & $[2]_6$ &  10/3  & 6088.49 $\pm$ 55.00 & 6217.45  \\
  & 5/2 & 0  & -4/3 & $[2]_6$ &  10/3 & 6203.49 $\pm$ 55.81 & 6250.80 \\
 \hline
$uubb\overline{b}$, $udbb\overline{b}$, $ddbb\overline{b}$ & 1/2 & 0 & 2/3 & $[2]_6$ & 10/3 &  16109.80 $\pm$ 103.30 & 15869.30\\ 
    & 1/2 & 1 & 2/3 & $[2]_6$ & 10/3 &  16173.80 $\pm$ 103.33 & \\ 
   & 3/2 & 0 & 2/3 & $[2]_6$ &  10/3 & 16178.80 $\pm$ 103.40 & 15882.50 & \\
   & 3/2 & 1 & 2/3 & $[2]_6$ &  10/3 & 16242.80 $\pm$ 103.43 &  & \\
  & 5/2 & 0 & 2/3 & $[2]_6$ & 10/3  & 16293.80 $\pm$ 103.83 & 15900.40 & \\
   & 5/2 & 1 & 2/3 & $[2]_6$ & 10/3  & 16357.80 $\pm$ 103.86 & & \\
   \hline
     $usbb\overline{b}$, $dsbb\overline{b}$ & 1/2 & 1/2 & -1/3 & $[2]_6$ & 10/3 & 16294.70 $\pm$ 103.31  & 16038.50 & \\
     & 3/2 & 1/2 & -1/3 & $[2]_6$ &  10/3 & 16363.70 $\pm$ 103.40 & 16049.60 &   \\
  & 5/2 & 1/2  & 4/3 & $[2]_6$ &  16/3 & 16748.70 $\pm$ 103.83 & 16061.30 \\
     \hline
    $ssbb\overline{b}$ & 1/2  & 0  & -4/3 & $[2]_6$ & 10/3 & 16415.70 $\pm$ 103.32 & 16211.00 & \\ 
 & 3/2 & 0 & -4/3 & $[2]_6$ &  10/3  & 16484.70 $\pm$ 103.41 &  16220.00 \\
  & 5/2 & 0  & -4/3 & $[2]_6$ &  10/3 & 16599.70 $\pm$ 103.85 & 16229.60\\
    \hline
     \hline
       \end{tabular}
    \label{tab:mass1}
\end{table*}

 \begin{table*}[ht!]
        \centering
         \caption{Table for masses of triply heavy $q\overline{q}ccb$ or $q\overline{q}bbc$ and $qqcc\Bar{b}$ or $qqbb\Bar{c}$ - type pentaquarks in octet and sextet configuration of SU(3) flavor representation by using the extension of Gursey-Radicati mass formula and the effective mass scheme. All masses are in the units of MeV.}
          \begin{tabular}{cccccccccc}
      \hline
      \hline
       \textcolor{red}{$q\overline{q}QQQ$ - type Pentaquarks} & & &  &  &  & & \\
      \hline
   Quark Content &  S &  I &  Y & $SU(3)_f$ & $C_2(SU(3)_f)$  & GR mass formula & Effective mass Scheme & Ref. \cite{EPJC2} & Ref. \cite{Reference}\\
   \hline
$u\overline{u}ccb$, $u\overline{d}ccb$, $d\overline{d}ccb$ & 1/2 & 0 & 0 & $[21]_8$ & 3 &  9270.55 $\pm$ 50.26 & 9100.32 & 9130 & 9203\\ 
    & 1/2 & 1 & 0 & $[21]_8$ & 3 &  9334.55 $\pm$ 50.33 &  & 9350 & 9324\\ 
   & 3/2 & 0 & 0 & $[21]_8$ &  3 & 9339.55 $\pm$ 50.45 & 9114.64 & 9170 & 9253\\
   & 3/2 & 1 & 0 & $[21]_8$ &  3 & 9403.55 $\pm$ 50.52 &  & 9315 & 9313\\
  & 5/2 & 0 & 0 & $[21]_8$ & 3  & 9454.55 $\pm$ 51.34 & 9248.65 & 9256 & 9126\\
   & 5/2 & 1 & 0 & $[21]_8$ & 3  & 9518.55 $\pm$ 51.40 & &  & 9281\\
     \hline
     $u\overline{s}ccb$, $d \overline{s}ccb$ & 1/2 & 1/2 & 1 & $[21]_8$ & 3 & 9128.25 $\pm$ 50.28  & 9296.93 &  9311\\ 
     & 3/2 & 1/2 & 1 & $[21]_8$ &  3 & 9197.25 $\pm$ 50.47 & 9321.14 & 9353\\
      & 5/2 & 1/2  & 1 & $[21]_8$ &  3 & 9312.25 $\pm$ 51.36 & 9410.90 & 9444\\
     \hline
 $s\overline{u}ccb$, $s \overline{d}ccb$ & 1/2 & 1/2 & -1 & $[21]_8$ & 3 & 9444.85 $\pm$ 50.28  & 9296.93 & 9309\\ 
     & 3/2 & 1/2 & -1 & $[21]_8$ &  3 & 9513.85 $\pm$ 50.47 & 9321.14 & 9355\\
      & 5/2 & 1/2  & -1 & $[21]_8$ &  3 & 9628.85 $\pm$ 51.36 & 9410.90 & 9445\\
    \hline
    $s\overline{s}ccb$ & 1/2 & 0 & 0 & $[21]_8$ & 3 & 9270.55 $\pm$ 50.26 & 9490.23  & 9488\\ 
     & 3/2 & 0 & 0 & $[21]_8$ &  3 & 9339.55 $\pm$ 50.45 &  9506.45 & 9539\\
      & 5/2 & 0 & 0 & $[21]_8$ &  3 & 9454.55 $\pm$ 51.34 & 9576.46 & 9632\\
     \hline
    $u\overline{u}bbc$, $u \overline{d}bbc$, $d \overline{d}bbc$  & 1/2 & 0 & 0 & $[21]_8$ & 3 & 12736.00 $\pm$ 71.29 & 12448.80 & 12439 & 12647\\ 
    & 1/2 & 1 & 0 & $[21]_8$ & 3 &  12800.00 $\pm$ 71.34 &  & 12659 & 12685\\ 
   & 3/2 & 0 & 0 & $[21]_8$ &  3 & 12805.00 $\pm$ 71.43 & 12450.60 & 12459 & 12462\\
   & 3/2 & 1 & 0 & $[21]_8$ &  3 & 12869.00 $\pm$ 71.47 &  & 12631 & 12645\\
  & 5/2 & 0 & 0 & $[21]_8$ & 3  & 12920.00 $\pm$ 72.05 & 12572.20 & 12584 & 12463\\
   & 5/2 & 1 & 0 & $[21]_8$ & 3  & 12939.00 $\pm$ 72.10 &  & & 12619\\
   \hline
$u\overline{s}bbc$, $d \overline{s}bbc$ & 1/2 & 1/2 & 1 & $[21]_8$ & 3 & 12593.70 $\pm$ 71.30  & 12642.60 & 12621\\ 
& 3/2 & 1/2 & 1 & $[21]_8$ &  3 & 12662.70 $\pm$ 71.44 & 12654.40 & 12641\\
& 5/2 & 1/2  & 1 & $[21]_8$ &  3 & 12777.70 $\pm$ 72.07 & 12735.80 & 12771\\
     \hline
$s\overline{u}bbc$, $s \overline{d}bbc$  & 1/2 & 1/2 & -1 & $[21]_8$ & 3 & 12910.30 $\pm$ 71.30  & 12642.60 & 12623\\ 
     & 3/2 & 1/2 & -1 & $[21]_8$ &  3 & 12979.30 $\pm$ 71.44 & 12654.40 & 12642\\
      & 5/2 & 1/2  & -1 & $[21]_8$ &  3 &  
 13094.30 $\pm$ 72.07 & 12735.80 & 12770\\
     \hline
$s\overline{s}bbc$ & 1/2 & 0 & 0 & $[21]_8$ & 3 & 12736.00 $\pm$ 71.29  & 12833.20 & 12804\\ 
& 3/2 & 0 & 0 & $[21]_8$ &  3 & 12805.00 $\pm$ 71.43 & 12841.10 & 12823\\
 & 5/2 & 0  & 0 & $[21]_8$ &  3 & 12920.00 $\pm$ 72.05 & 12902.70 & 12958\\
     \hline
       \textcolor{red}{$qqQQ\overline{Q}$ - type Pentaquarks} &  &  &  &  &  &\\
     \hline
    $uucc\overline{b}$, $udcc\overline{b}$, $ddcc\overline{b}$  & 1/2 & 0 & 2/3 & $[2]_6$ & 10/3 &  9178.96 $\pm$ 50.30 & 9191.71\\ 
    & 1/2 & 1 & 2/3 & $[2]_6$ & 10/3 & 9242.96 $\pm$ 50.36 & \\ 
   & 3/2 & 0 & 2/3 & $[2]_6$ &  10/3 & 9247.96 $\pm$ 50.50 & 9233.06 & \\
   & 3/2 & 1 & 2/3 & $[2]_6$ &  10/3 & 9311.96 $\pm$ 50.56 &  & \\
  & 5/2 & 0 & 2/3 & $[2]_6$ & 10/3  & 9362.96 $\pm$ 51.38 & 9248.66 & \\
   & 5/2 & 1 & 2/3 & $[2]_6$ & 10/3  & 9426.96 $\pm$ 51.44 &  & \\
     \hline
$uscc\overline{b}$, $dscc\overline{b}$  & 1/2 & 1/2 & -1/3 & $[2]_6$ & 10/3 & 9363.93 $\pm$ 50.31  & 9362.31 & \\ 
& 3/2 & 1/2 & -1/3 & $[2]_6$ &  10/3 & 9432.93 $\pm$ 50.50 & 9397.43 & \\
  & 5/2 & 1/2  & -1/3 & $[2]_6$ &  10/3 & 9547.93 $\pm$ 51.38 & 9410.90 & \\
     \hline
    $sscc\overline{b}$ & 1/2  & 0  & -4/3 & $[2]_6$ & 10/3 & 9489.89 $\pm$ 50.33 & 9536.22 & \\ 
     & 3/2 & 0 & -4/3 & $[2]_6$ &  10/3  & 9553.89 $\pm$ 50.52 & 9565.11  \\
  & 5/2 & 0  & -4/3 & $[2]_6$ &  10/3 & 9668.89 $\pm$ 51.40 & 9576.47 \\
 \hline
$uubb\overline{c}$, $udbb\overline{c}$, $ddbb\overline{c}$ & 1/2 & 0 & 2/3 & $[2]_6$ & 10/3 &  12644.40 $\pm$ 71.32 & 12518.80\\ 
    & 1/2 & 1 & 2/3 & $[2]_6$ & 10/3 &  12708.40 $\pm$ 71.36 & \\ 
   & 3/2 & 0 & 2/3 & $[2]_6$ &  10/3 & 12713.40 $\pm$ 71.46 & 12531.70 & \\
   & 3/2 & 1 & 2/3 & $[2]_6$ &  10/3 & 12777.40 $\pm$ 71.50 &  & \\
  & 5/2 & 0 & 2/3 & $[2]_6$ & 10/3  & 12828.40 $\pm$ 72.08 & 12572.20 & \\
   & 5/2 & 1 & 2/3 & $[2]_6$ & 10/3  & 12892.40 $\pm$ 72.13 & & \\
   \hline
     $usbb\overline{c}$, $dsbb\overline{c}$ & 1/2 & 1/2 & -1/3 & $[2]_6$ & 10/3 & 12829.30 $\pm$ 71.32  & 12690.70 & \\
     & 3/2 & 1/2 & -1/3 & $[2]_6$ &  10/3 & 12898.30 $\pm$ 71.46 & 12701.50 &   \\
  & 5/2 & 1/2  & 4/3 & $[2]_6$ &  16/3 & 13013.30 $\pm$ 72.09 & 12735.80 \\
     \hline
    $ssbb\overline{c}$ & 1/2  & 0  & -4/3 & $[2]_6$ & 10/3 & 12950.30 $\pm$ 71.34 & 12886.20 & \\ 
 & 3/2 & 0 & -4/3 & $[2]_6$ &  10/3  & 13019.30 $\pm$ 71.47 &  12874.70 \\
  & 5/2 & 0  & -4/3 & $[2]_6$ &  10/3 & 13134.30 $\pm$ 72.10 & 12902.70\\
    \hline
     \hline
       \end{tabular}
    \label{tab:mass3}
\end{table*}

\subsection{Screened Charge Scheme}
The screened charge scheme, derived from Quantum Chromodynamics (QCD), acknowledges the dynamic nature of the strong nuclear force and its interaction with quarks and gluons. In the context of triply heavy pentaquarks, consisting of three heavy quarks (e.g., charm or bottom quarks) and two light quarks (up, down, or strange), multiple heavy quarks introduce unique challenges and opportunities for understanding their properties. By applying the screened charge scheme to triply heavy pentaquarks, one can investigate how the strong interaction between heavy quarks and surrounding quark-gluon or nuclear medium affects the particle properties, such as their mass, decay modes, and production rates. Furthermore, the screening effect may alter heavy quark binding energies and spatial distributions within pentaquark structures, offering insights into their internal structure and dynamics. Therefore, the effective charge of the quark '$q$' in pentaquark ($q$, $r$, $v$, $x$, $y$) can be written as: 
\begin{equation}
    e_q^P = e_q + \alpha_{qr} e_r + \alpha_{qv} e_v + \alpha_{qx} e_x + \alpha_{qy} e_y
    \label{screen1}
    \end{equation}
Similarly, we can write this for other quarks (anti-quarks) inside the pentaquarks \cite{sharma2024}. Where $e_q$ stands for the charge of the quark (anti-quark). By considering the isospin symmetry:
\begin{align}
    \alpha_{uu} = \alpha_{ud} = \alpha_{dd} = \alpha_1 \\ \nonumber
    \alpha_{us} = \alpha_{ds} = \beta_1 \\ \nonumber
    \alpha_{ss} = \beta_2
\end{align}
 for charm states,
 \begin{align}
     \alpha_{uc} = \alpha_{dc} = \beta_3 \\ \nonumber
     \alpha_{sc} = \alpha_2, \hspace{0.3cm} \alpha_{cc} = \alpha_3
 \end{align}
Similarly, for the bottom sector,
 \begin{align}
     \alpha_{ub} = \alpha_{db} = \beta_4, \hspace{0.3cm} \alpha_{sb} = \alpha_4 \\ \nonumber
     \alpha_{cb} = \beta_5, \hspace{0.3cm} \alpha_{bb} = \alpha_5
 \end{align}
with the help of SU(3) symmetry, then we can reduce these parameters as:
\begin{equation}
    \alpha_1 = \beta_1 = \beta_2
\end{equation}
To calculate the screening parameter $\alpha_{ij}$, we used the Ansatz formalism, 
\begin{equation}
    \alpha_{ij} = \mid{\frac{m_i - m_j}{m_i + m_j}}\mid \times \delta
\end{equation}
here, $m_i$ and $m_j$ are the respective quark masses and $\delta$ = 0.81 \cite{Bains}. The value of parameters helps us predict the magnetic moments of triply heavy pentaquark states. By putting the values of parameters in effective charge equations and by introducing the magnetic moment operator:
\begin{equation}
    \mu = \sum_i \frac{e_i^P}{2 m_i^{eff}} \sigma_i
\end{equation}
Now, the magnetic moment operator consists of two parts:
 \begin{equation}
     \Vec{\mu} = \Vec{\mu}_{spin} + \Vec{\mu}_{orbit}
 \end{equation}
which can be written as:
\begin{equation}
    \Vec{\mu} = \hspace{0.3cm} \sum_i \mu_i (2 \Vec{s_i} + \Vec{l_i}) = \hspace{0.3cm} \sum_i \mu_i(2\Vec{s_i}) = \hspace{0.3cm} \sum_i \mu_i(\Vec{\sigma_i})
\label{Magnetic moment}
\end{equation}
Due to the absence of orbital excitations, the magnetic moment depends on the spin part. Thus, by calculating the expectation value of Eq.\eqref{Magnetic moment} using the spin-flavor wavefunction of triply heavy pentaquarks, magnetic moments can be defined as:
\begin{equation}
    \mu = \bra{\Psi_{sf}}\Vec{\mu}\ket{\Psi_{sf}}
    \label{magnetic123}
\end{equation}
By substituting the spin-flavor wavefunction for both $qqQQ\Bar{Q}$ and $q\Bar{q}QQQ$ type triply heavy Pentaquarks in equation \eqref{magnetic123}, we estimated the magnetic moments using the effective mass and screened charge scheme, which is reported in Table \ref{tab: magnetic moment1} and \ref{tab: magnetic moments}. Expressions for magnetic moments for $J^P$ values $5/2^-$, $3/2^-$, and $1/2^-$ are written in Table \ref{tab: expressions}. We used $\chi_1^P$,  $\chi_4^P$, and  $\chi_7^P$ symmetries for $J^P$ equals to $5/2^-$, $3/2^-$, and $1/2^-$ respectively. These symmetries provide fewer deviations in magnetic moment assignments by using the methodology of effective mass and screened charge schemes. The next subsection discussed the results of $qqQQ\Bar{Q}$ and $q\Bar{q}QQQ$ type triply heavy Pentaquarks.

\begin{table}[t]
    \centering
      \caption{Magnetic moments expressions for pentaquarks with $J^P$ values $5/2^-$, $3/2^-$ and $1/2^-$ respectively (in $\mu_N$) \cite{young}.}
    \begin{tabular}{cc}
    \hline
    \hline
     Basis  & \hspace{0.3cm} Magnetic Moment Expression($\mu$) \\
       \hline
        $\chi_1^P$  & \hspace{0.3cm}  $\mu_1$ + $\mu_2$ + $\mu_3$ + $\mu_4$ + $\mu_5$ \\
        \\
        $\chi_2^P$ & \hspace{0.3cm} $\frac{9}{10}$ ($\mu_1$ + $\mu_2$ + $\mu_3$ + $\mu_4$) -$\frac{3}{5}$ $\mu_5$\\
        \\
        $\chi_3^P$ & \hspace{0.3cm} $\frac{5}{6}$ ($\mu_1$ + $\mu_2$ + $\mu_3$) -$\frac{1}{2}$ $\mu_4$ + $\mu_5$\\
        \\
       $\chi_4^P$ & \hspace{0.3cm} $\frac{2}{3}$ ($\mu_1$ + $\mu_2$) - $\frac{1}{3}$ $\mu_3$ + $\frac{5}{6}$ ($\mu_4$ + $\mu_5$)\\
       \\
       
       $\chi_5^P$ & \hspace{0.3cm} $\mu_3$ + $\mu_4$ + $\mu_5$ \\
       \\
        $\chi_6^P$ & \hspace{0.3cm} $\frac{5}{6}$ ($\mu_1$ + $\mu_2$ + $\mu_3$) -$\frac{1}{3}$ ($\mu_4$ + $\mu_5$) \\
        \\
        $\chi_7^P$ & \hspace{0.3cm} $\frac{4}{9}$ ($\mu_1$ + $\mu_2$) - $\frac{2}{9}$ $\mu_3$ + $\frac{2}{3}$$\mu_4$ -  $\frac{1}{3}$ $\mu_5$   \\
        \\
       $\chi_8^P$ & \hspace{0.3cm}  $\frac{2}{3}$ ($\mu_3$ + $\mu_4$) - $\frac{1}{3}$ $\mu_5$ \\
       \\
        $\chi_9^P$ & \hspace{0.3cm} $\mu_5$  \\
        \\
        $\chi_{10}^P$ & \hspace{0.3cm} $\mu_5$ \\
        
        \hline
        \hline
    \end{tabular}
  \label{tab: expressions}
\end{table}

\begin{table*}[]
    \centering
    \caption{Magnetic moments of $qqcc\Bar{c}$, $qqbb\Bar{b}$  and $q\Bar{q}ccc$, $q\Bar{q}bbb$ - type triply heavy Pentaquarks with all possible $J^P$ values using the eff. mass scheme, screened charge scheme, and both schemes together.}
    \renewcommand{\arraystretch}{1.0}
 \tabcolsep 0.01cm
    \begin{tabular}{|c|ccc|ccc|ccc|}
    \hline
   
      Quark Contents  & &  Eff. Mass Scheme & & \multicolumn{2}{c}{\hspace{0.2cm} Screened Charge Scheme} & & \multicolumn{2}{c}{\hspace{0.2cm} Eff. mass + Screen. Charge Scheme} &\\
      \hline
        & $1/2^-$  & $3/2^-$ &  $5/2^-$ & $1/2^-$ & $3/2^-$ & $5/2^-$ & $1/2^-$ & $3/2^-$ & $5/2^-$ \\
        \hline

$uucc\Bar{c}$  & 0.61 & 2.08 & 2.21 & 0.99 & 3.91 & 6.79 & 0.99 & 3.76 & 6.45\\
$udcc\Bar{c}$  & 0.61 & 0.42 & 1.32 & 0.70 & 1.79 & 3.33 & 0.70 & 1.72 & 3.16\\
$ddcc\Bar{c}$  & 0.61 & -1.23 & -1.10 & 0.41 & -0.31 & -0.12 & 0.41 & -0.32 & -0.12  \\
$uscc\Bar{c}$  & 0.61 & 0.61 & 1.34 & -2.21 & 1.76 & 3.30 & 0.72 & 1.71 & 3.21 \\
$dscc\Bar{c}$  & 0.61 & -1.06 & -1.19 & 0.43 & -0.53 & -0.43 & 0.43 & -0.53 & -0.50  \\
$sscc\Bar{c}$  & 0.61 & -0.88 & -0.77 & 0.45 & -0.45 & -0.28 & 0.45 & -0.45 & -0.27 \\
\hline
 $uubb\Bar{b}$  & -0.10 & 2.23 & 1.54 & 0.07 & 1.75 & 2.71 & 0.07 & 1.69 & 2.60\\
$udbb\Bar{b}$  & -0.10 & 0.57 & 1.69 & -0.06 & -0.14 & -0.27 & -0.06 & -0.13 & -0.25\\
$ddbb\Bar{b}$  & -0.10 & -1.08 & -0.77 & -0.18 & -2.04 & -3.25 & -0.18 & -1.97 & -3.11  \\
$usbb\Bar{b}$  & -0.10 & 0.76 & 1.70 & -0.05 & 0.22 & 0.29 & -0.05 & 0.21 & 0.27 \\
$dsbb\Bar{b}$  & -0.10 & -0.92 & -0.79 & -0.18 & -1.86 & -2.96 & -0.18 & -1.81 & -2.88  \\
$ssbb\Bar{b}$  & -0.10 & -0.74 & -0.35 & -0.18 & -1.36 & -2.22 & -0.18 & -1.35 & -2.19 \\
\hline
$u\Bar{u}ccc$  & 2.15 & 0.23 & 0.49 & 2.88 & 4.91 & 5.64  & 3.15 & 5.26 & 5.35 \\
$u\Bar{d}ccc$  & 1.19 & 2.72 & 2.92 & 2.21 & 7.36 & 10.01 & 2.39 & 8.05 & 9.49 \\
$d\Bar{d}ccc$  & -0.70 & 0.43 & 0.49 & 0.28 & 4.91 & 6.55 & 0.29 & 5.47 & 6.20  \\
$u\Bar{s}ccc$ & 1.29 & 2.36  & 2.70 & 2.74 & 6.32 & 8.73 & 2.98 & 6.55 & 8.42 \\
$d\Bar{s}ccc$  & -0.58 & 0.13 & 0.24 & 0.91 & 3.64 & 4.99 & 1.00 & 3.79 &  4.83 \\
$s\Bar{s}ccc$  & -0.35 & 0.13  & 0.498 & 0.15 & 2.80  & -0.80 & 0.14 & 3.30 &  3.95 \\
$s\Bar{u}ccc$  & -1.99 & -1.88 & -3.28 & -0.36 & -0.84 &  -1.70 & -0.54 & -0.77 & -1.55\\
$s\Bar{d}ccc$  & -1.94 & -0.14 & -0.74 & -0.84 & 0.80 & 1.28 & -1.02 & 0.76 & 1.18\\
\hline
$u\Bar{u}bbb$  & 1.79 & -0.10 & -0.06 & 1.08 & -3.07 & -3.79  & 1.15 & -3.14 & -3.62 \\
$u\Bar{d}bbb$  & 0.87 & 2.11 & 2.41 & 0.30 & -0.77 & -0.81 & 0.32 & -0.79 & -0.76 \\
$d\Bar{d}bbb$  & -0.96 & -0.04 & -0.06 & -1.50 & -3.07 & -3.79 & -1.60 & -3.08 & -3.62  \\
$u\Bar{s}bbb$  & 0.96 & 1.95  & 2.17 & 0.20 & -0.19  & -0.12 & 0.21 & -0.20 &  -0.11 \\
$d\Bar{s}bbb$ & -0.84 & -0.31  & -0.32 & -0.73 & -2.71 & -3.37 & -1.59 & -2.84 & -3.27 \\
$s\Bar{s}bbb$  & -0.63 & -0.05 & -0.05 & -1.00 & -1.95 & -2.45 & -1.02 & -2.00 &  -2.42 \\
$s\Bar{u}bbb$  & -1.99 & -1.88 & -3.28 & -0.36 & -0.84 &  -1.70 & -0.54 & -0.77 & -1.55\\
$s\Bar{d}bbb$  & -1.94 & -0.14 & -0.74 & -0.84 & 0.80 & 1.28 & -1.02 & 0.76 & 1.18\\
\hline
    \end{tabular}
    \label{tab: magnetic moment1}
\end{table*}

\begin{table*}[]
    \centering
    \caption{Magnetic moments of $qqcc\Bar{b}$, $qqcc\Bar{b}$  and $q\Bar{q}ccb$, $q\Bar{q}bbc$ - type  Pentaquarks with possible $J^P$ values using the formalism of eff. mass scheme, screened charge scheme, and both schemes together.}
    \renewcommand{\arraystretch}{1.0}
 \tabcolsep 0.01cm
    \begin{tabular}{|c|ccc|ccc|ccc|}
    \hline
   
Quark Content  & &  Eff. Mass Scheme & & \multicolumn{2}{c}{\hspace{0.2cm} Screened Charge Scheme} & & \multicolumn{2}{c}{\hspace{0.2cm} Eff. mass + Screened Charge Scheme} &\\
      \hline
& $1/2^-$  & $3/2^-$ &  $5/2^-$ & $1/2^-$ & $3/2^-$ & $5/2^-$ & $1/2^-$ & $3/2^-$ & $5/2^-$ \\
        \hline

$uucc\Bar{b}$  & 1.93 & 2.42 & 4.04 & 4.68 & 6.22 & 10.28 & 4.56 & 5.94 & 8.82\\
$udcc\Bar{b}$  & 0.52 & 0.78 & 1.60 & 3.22 & 4.24 & 6.98 & 3.13 & 4.05 & 5.86\\
$ddcc\Bar{b}$  & -0.87 & -0.86 & -0.83 & 1.75 & 2.26 & 3.68 & 1.71 & 2.15 & 3.06  \\
$uscc\Bar{b}$  & 0.68 & 0.96 & 1.87 & 2.86 & 3.76 & 6.28 & 2.83 & 3.84 & 5.27 \\
$dscc\Bar{b}$  & -0.73 & -0.69 & -0.59 & 1.20 & 1.59 & 2.70 & 1.18 & 1.55 & 2.16 \\
$sscc\Bar{b}$  & -0.58 & -0.52 & -0.34 & 0.79 & 1.24 & 2.19 & 0.78 & 1.22 & 1.67 \\
\hline
 $uubb\Bar{c}$  & 1.99 & 1.90 & 2.78 & -0.37 & -0.58 & -0.77 & -0.36 & -0.57 & -0.84\\
$udbb\Bar{c}$  & 0.57 & 0.22 & 0.33 & -1.74 & -2.62 & -3.91 & -1.70 & -2.55 & -3.59\\
$ddbb\Bar{c}$  & -0.83 & -1.45 & -2.12 & -3.11 & -4.65 & -7.05 & -3.04 & -4.54 & -6.30  \\
$usbb\Bar{c}$  & 0.74 & 0.41 & 0.60 & -1.20 & -1.80 & -2.69 & -1.19 & -1.78 & -2.49 \\
$dsbb\Bar{c}$  & -0.68 & -1.28 & -1.88 & -2.61 & -4.02 & -6.19 & -2.58 & -3.97 & -5.50 \\
$ssbb\Bar{c}$  & -0.68 & -1.10 & -1.62 & -1.84 & -3.08 & -4.70 & -2.32 & -3.07 & -4.21 \\
\hline
$u\Bar{u}ccb$ & 0.56 & 0.51 & 0.67 & 1.49 & 2.40 & 5.64 & 1.86 & 2.55 & 5.35 \\
$u\Bar{d}ccb$ & -0.37 & 2.82 & 3.11 & -0.93 & 4.91 & 10.01 & -0.71 & 5.21 & 9.49 \\
$d\Bar{d}ccb$ & 0.56 & 0.51 & 0.43 & -0.56 & 2.40 & 6.55 & -0.35 & 2.55 & 6.20  \\
$d\Bar{u}ccb$  & 1.50 & -1.80 & 0.49 & -0.17 & -0.09 & 6.55 & -0.01 & -0.10 & 6.20  \\
$u\Bar{s}ccb$ & -0.25 & 2.52 & 2.70 & -0.76 & 4.40 & 8.73 & -0.49 & 4.56 & 8.42 \\
$d\Bar{s}ccb$  & 0.67 & 0.26 & 0.24 & -0.28 & 1.66 & 4.99 & -0.05 & 1.73 &  4.83 \\
$s\Bar{s}ccb$  & 0.56 & 0.51  & 0.498 & -0.14 & 1.37  & -0.80 & 1.39 & 3.30 &  3.95 \\
$s\Bar{u}ccb$  & 1.38 & -1.50 & -1.06 & 0.06 & -0.62 &  -1.70 & -0.65 & -0.77 & -1.55\\
$s\Bar{d}ccb$  & 0.45 & 0.75 & 2.38 & -0.41 & 2.11 & 1.28 & -0.18 & 2.18 & 1.18\\
\hline
$u\Bar{u}bbc$  & 0.20 & -0.20 & 0.24 & 0.35 & -0.58 & -0.39 & 0.37 & -0.61 & -1.64 \\
$u\Bar{d}bbc$  & -0.50 & 2.11 & 2.69 & 0.10 & 1.65 & 2.79 & 0.09 & 1.77 & 1.46 \\
$d\Bar{d}bbc$  & 0.09 & -0.20 & 0.24 & 0.35 & -0.58 & -0.39 & 0.22 & -0.61 & 0.27  \\
$d\Bar{u}bbc$  & 1.13 & -2.52 & -2.21 & 0.76 & -2.81 & -3.53 & 0.84 & -3.01 & -2.12  \\
$u\Bar{s}bbc$  & -0.60 & 1.56  & 2.44 & -0.09 & 1.72  & 2.78 & -0.12 & 1.54 & 1.43 \\
$d\Bar{s}bbc$ & 0.31 & 0.03  & -0.01 & 0.41 & -0.74 & -0.63 & 0.42 & -0.41 & -0.01 \\
$s\Bar{s}bbc$  & 0.20 & -0.20 & 0.24 & 0.34 & -0.53 & 0.24 & 0.35 & -0.54 & 0.13 \\
$s\Bar{u}bbc$  & 1.01 & -2.21 & -1.96 & 0.79 & -2.83 & -3.52 & 0.84 & -2.95 & -2.27\\
$s\Bar{d}bbc$  & 0.09 & 0.03 & 0.49 & 0.32 & -0.36 & -0.10 & 0.33 & -0.38 & 0.21\\

\hline
    \end{tabular}
    \label{tab: magnetic moments}
\end{table*}

\section{Results and Discussion}
In this work, we performed the phenomenological analysis of triply heavy pentaquarks with configurations $qqQQ\Bar{Q}$ and $q\Bar{q}QQQ$ ($q = u, d, s$ and $Q = c, b$). The possible configurations are $qqcc\Bar{c}$, $qqbb\Bar{b}$, $qqcc\Bar{b}$, $qqbb\Bar{c}$ for $qqQQ\Bar{Q}$ - type pentaquarks and  $q\Bar{q}ccc$, $q\Bar{q}bbb$, $q\Bar{q}ccb$, and $q\Bar{q}bbc$ for $q\Bar{q}QQQ$ - type pentaquarks. We used special unitary representation and Young tableau techniques to study the classification schemes for triply heavy pentaquarks. Further, mass spectra for triply heavy pentaquarks have been calculated using the formalism of the extended GR mass formula and the effective mass scheme. Moreover, the magnetic moment assignments for triply heavy pentaquarks are estimated using the methodologies of effective mass and screened charge schemes. Now, we will analyze the possible triply heavy pentaquark configurations in detail as follows: 

\subsection{$qqQQ\Bar{Q}$- Type Pentaquarks}
To classify the $qqQQ\Bar{Q}$ - type pentaquarks, we used the SU(3) flavor representation. The allowed flavor multiplets are anti-triplet and a sextet. We classified the $qqQQ\Bar{Q}$ - type pentaquarks into sextet configuration and studied the masses and magnetic moments. The possible quark contents for $qqQQ\Bar{Q}$ - type pentaquarks are $uuQQ\Bar{Q}$, $udQQ\Bar{Q}$, $ddQQ\Bar{Q}$, which are isospin partners with possible isospin values as 0 and 1. Also, $usQQ\Bar{Q}$ and $dsQQ\Bar{Q}$ are isospin partners with isospin 1/2. Further, $ssQQ\Bar{Q}$ is an isospin singlet state. Here, $qqQQ\Bar{Q}$- type pentaquarks consist of all possible configurations that contain triply charm, triply bottom, and mixed combinations. Masses of $qqQQ\Bar{Q}$ - type pentaquarks using the extended version of the GR mass formula and the effective mass scheme are reported in Tables \ref{tab:mass1} and \ref{tab:mass3} and their magnetic moment assignments using the formalism of effective mass and screened charge schemes are written in Tables \ref{tab: magnetic moment1} and \ref{tab: magnetic moments}. We compare our prediction with Ref. \cite{EPJC1} and Ref. \cite{EPJC2}, which shows quite good agreement and may be helpful for experimental studies in the future.

\subsection{$q\Bar{q}QQQ$- Type Pentaquarks}
We classified the $q\overline{q}QQQ$ - type pentaquarks into the octet configuration of the SU(3) flavor representation. $u\overline{u}QQQ$, $u\overline{d}QQQ$, and $d\overline{d}QQQ$ forms isospin triplet and thus have same mass. In a similar manner,  $u\overline{s}QQQ$, $d\overline{s}QQQ$, $s\overline{u}QQQ$, and $s\overline{d}QQQ$ are isospin-1/2 states and exhibits same mass. Further,  $s\overline{s}QQQ$ is an isospin-singlet state with no isospin partners.  Masses of $q\overline{q}QQQ$ - type pentaquarks are reported in Tables \ref{tab:mass1} and \ref{tab:mass3}, and their magnetic moment assignments using effective mass and screened charge schemes are reported in Tables \ref{tab: magnetic moment1} and \ref{tab: magnetic moments}. We compared our analysis for masses with the available theoretical data, which aligns well with us.

\section{Summary}
In this study, we conducted a spectroscopic analysis of the masses and magnetic moments of triply heavy pentaquarks. Employing a classification system based on various multiplets within the SU(3) flavor representation. This allowed us to categorize triply heavy pentaquarks according to their distinct flavor symmetries, facilitating a comprehensive analysis of their properties. We classified the $qqQQ\Bar{Q}$ and $q\Bar{q}QQQ$ into sextet and octet configurations, respectively. We used the formalism of the extended GR mass formula and the effective mass scheme to calculate the masses of triply heavy pentaquarks. Also, to calculate the magnetic moment assignments, we used the effective mass and screened charge schemes. Furthermore, to validate the accuracy and vitality of our calculations, we conducted thorough comparisons with existing theoretical data. This comparative analysis enables us to assess the consistency of our results with prior theoretical predictions. In conclusion, this work has thoroughly analyzed triply heavy pentaquarks through theoretical modeling and computational analysis. Our findings (masses and magnetic moments) underscore the importance of triply heavy pentaquarks in advancing our understanding of the exotic hadron field. While limitations such as the absence of experimental data may have constrained the scope of our study, our research offers valuable contributions to the theoretical framework of triply heavy pentaquarks and provides a foundation for future investigations.

\nocite{*}

\bibliography{triply}

\end{document}